\documentclass[linenumbers]{elsarticle}
\usepackage[latin1]{inputenc}
\usepackage[english]{babel}
\usepackage{graphicx}
\usepackage{color}
\usepackage{epsfig}
\usepackage{amssymb}
\usepackage{amsmath}
\usepackage{a4wide}
\usepackage{verbatim}
\usepackage{fancyhdr}
\usepackage{subfigure}
\usepackage{float}
\usepackage{fixltx2e}
\usepackage[ps2pdf]{thumbpdf}
\usepackage[ps2pdf]{hyperref}
\usepackage[numbers]{natbib}
\usepackage{setspace}

\begin{document}

\newcommand{\Angst}{$\mathring{\mathrm{A}}$}
\newcommand{\DEG}{$^\circ$}
\newcommand{\sq}{$^\mathrm{2}$\,} 
\begin{frontmatter}

\title{Bi-spectral beam extraction in combination with a focusing feeder}
\author[hzb,workpackage]{C. Zendler\corref{cor1}}
\ead{carolin.zendler@helmholtz-berlin.de, ph: +49-30-8062-43070, fax: +49-30-8062-43094}
\author[hzb,workpackage]{K. Lieutenant}
\author[hzb,workpackage]{D. Nekrassov}
\author[hzb,workpackage]{L. D. Cussen}
\author[ess]{M. Strobl}
\address[hzb]{Helmholtz-Zentrum Berlin f{\"u}r Materialien und Energie GmbH, Hahn-Meitner-Platz 1, 14109 Berlin, Germany}
\address[workpackage]{ESS Design Update Programme, Germany}
\address[ess]{European Spallation Source ESS AB, Box 176, 22100 Lund, Sweden}
\cortext[cor1]{Correspondig author}

\begin{abstract}
Bi-spectral beam extraction combines neutrons from two different kind of moderators into one beamline, expanding the spectral range and thereby the utilization of an instrument. This idea can be realized by a mirror that reflects long wavelength neutrons from an off-axis colder moderator into a neutron guide aligned with another moderator emitting neutrons with shorter wavelengths which will be transmitted through the mirror. The mirror used in such systems is typically several meters long, which is a severe disadvantage because it reduces the possible length of a focusing device in design concepts requiring a narrow beam at a short distance from the source, as used in many instruments under development for the planned European Spallation Source (ESS). We propose a shortened extraction system consisting of several mirrors, and show that such an extraction system is better suited for combination with a feeder in an eye of the needle design, illustrated here in the context of a possible ESS imaging beamline. 
\end{abstract}

\end{frontmatter}
\section{Introduction: concept of bi-spectral beam extraction}

\begin{figure}[tb!]
\centering
\subfigure[Principle of bi-spectral extraction with a mirror of length 2.34\,m. Arrows represent example trajectories of neutrons.\label{f_Skizze1mirr}]{\includegraphics[width=.45\linewidth]{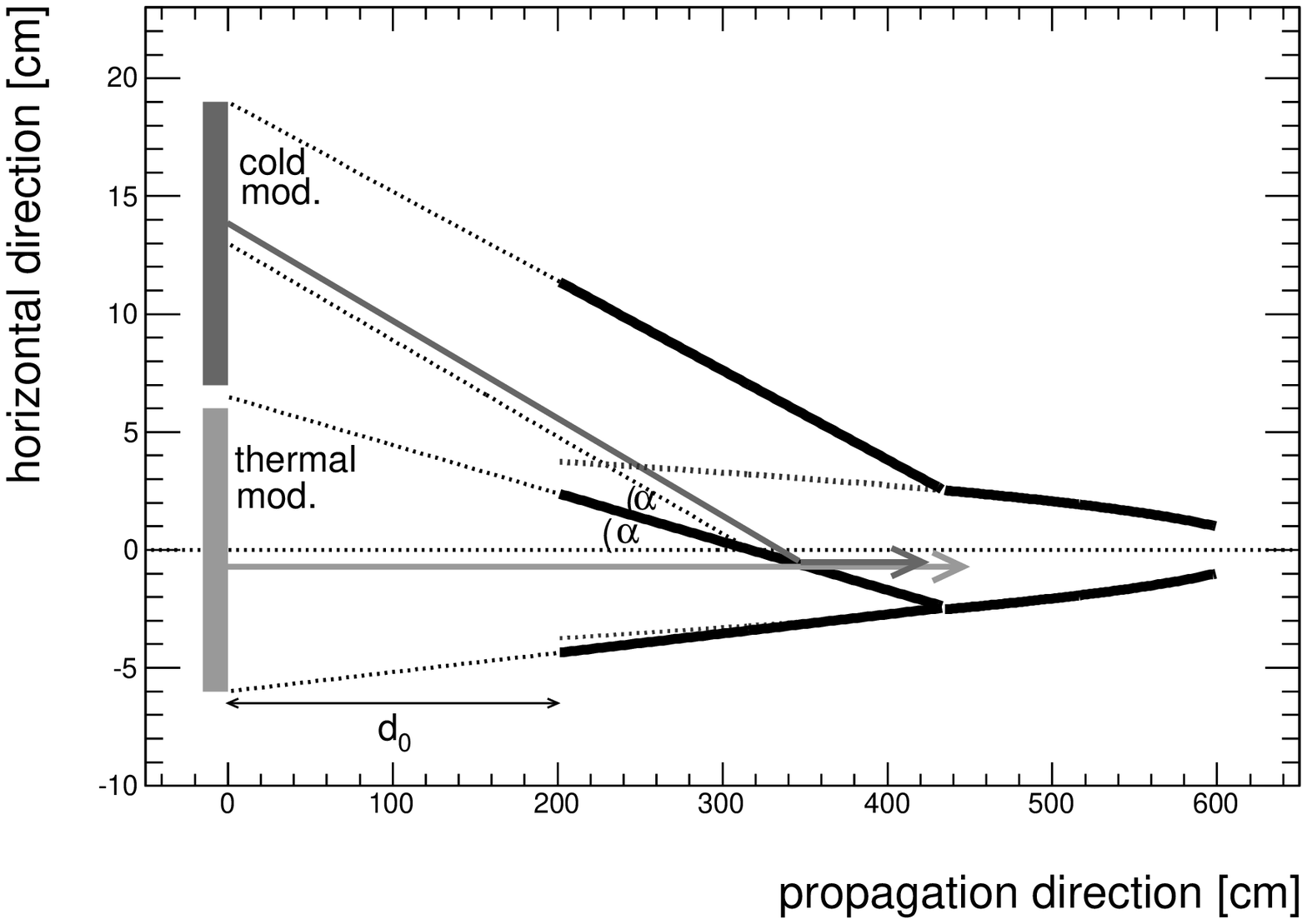}}
\subfigure[Extraction system with 4 mirrors of 1\,m length.\label{f_Skizze4mirr}]{\includegraphics[width=.45\linewidth]{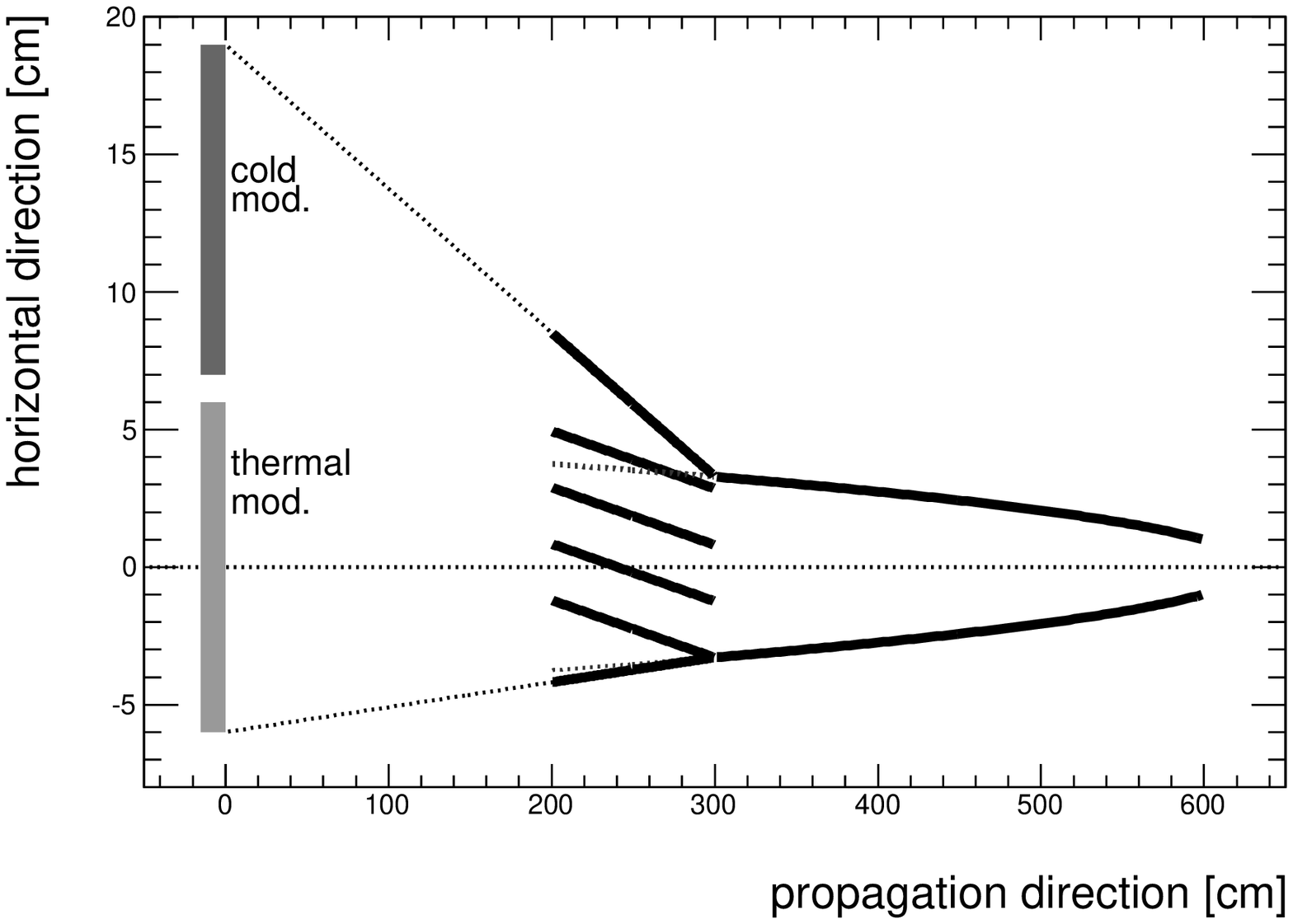}}
\caption{Schematic drawing of extraction systems, top view. Dotted lines are projections or indicate the original feeder shape.}
\label{f_Skizzen1}
\end{figure}

For instruments at the European Spallation Source (ESS)~\cite{ESSWeb}, cold and thermal moderators with surface areas of 12$\times$12\,cm\sq\,each are planned to be available side by side. In order to transport neutrons from both these moderators into one beamline, a mirror can be used such that thermal neutrons pass through the mirror into a following guide which is aligned with the thermal moderator, while cold neutrons are reflected into the guide as illustrated in figure~\ref{f_Skizze1mirr}. In comparison to aligning the beamline with the center between moderators, such a mirror design is optimized to extract the maximum possible neutron flux from both moderators. This concept was first proposed by Mezei and Russina~\cite{ExtrSystMezeiPatent1,ExtrSystMezeiPatent2,Mezei} and has been studied with a following elliptic guide by Jacobsen et al.~\cite{ExtrSystHenrikKlaus}. It has been implemented once already at the EXED~\cite{ExedMC} beamline at HZB.

The mirror is pointing towards the separation between the moderators and is inclined at an angle $\alpha$ to the guide axis which is the same as the incident angle of neutrons emerging from the center of the cold moderator and hitting the center of the mirror. The length of the mirror is designed such that $\alpha$ is the critical angle for neutrons with the so-called cross-over wavelength $\lambda_{c}$, the wavelength for which the intensities from the cold and thermal moderators are equal. According to current ESS plans, $\lambda_{c}$ will be at 2.35\,\Angst. Thus $\alpha=0.1m\lambda_c$ (in deg if $\lambda$ in \Angst) is solely defined by the chosen mirror coating $m$, where the $m$-number represents the critical angle for total reflection in units of the critical angle of nickel. For $m$\,=\,5, $\alpha$\,=\,1.175\DEG, and with a 1\,cm gap between the moderators assumed, a mirror length of $L$\,=\,2.34\,m follows from $L_S=2 \cdot \left( D /(2\tan{\alpha}) - d_0 \right)$, where $D$ is the distance between the two moderators' centers, $d_0$ is the distance of the extraction system to the source (which has to be at least 2\,m in case of ESS, see figure~\ref{f_Skizzen1}), and $L_S$ is the length of the mirror projected on the guide axis. A lower mirror $m$-number requires longer mirrors. 
In figure~\ref{f_Skizzen1}, additional guide walls surrounding the extraction mirror are placed such that no direct path between source and guide remainder is blocked.

At ESS, shielding restricts the shortest possible distance from source to first chopper to 6\,m, while a guide tube can be placed starting from 2\,m. Many instrument designs therefore incorporate a 4\,m long focusing guide (``feeder'') to produce a narrow beam at 6\,m. If a 2.34\,m long extraction mirror must be inserted into this feeder, at least one guide wall must be removed to avoid blocking neutrons from the cold source before they reach the mirror, leaving less than 2\,m to focus the beam.

Figure~\ref{f_Skizze4mirr} is a schematic drawing of an alternative extraction system which consists of several mirrors, arranged such that neutrons emerging from the thermal moderator parallel to the guide axis cross only one mirror. This design has the advantage that the mirror system is shorter, and thus a larger fraction of the feeder can be used. Hence this approach should be particularly useful in an eye-of-the-needle concept with a chopper window at around 6\,m at ESS, or for any other beamline that requires a focal point at a short distance from the source. The alternative of one long mirror extracting neutrons from two moderators into one long (elliptical) guide with no ``eye of the needle'' is described in \cite{ExtrSystHenrikKlaus}.

\section{Performance of extraction systems with several mirrors}

\subsection{Simulation details}

\begin{figure}[tb!]
\centering
\includegraphics[width=.9\linewidth]{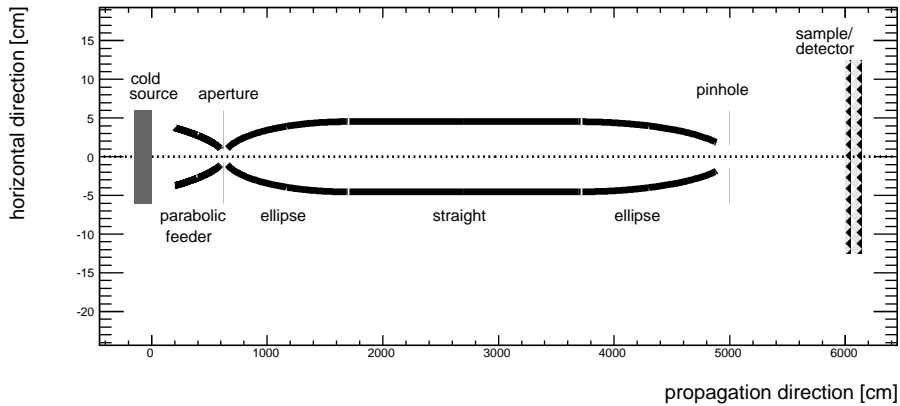}
\caption{Schematic drawing of possible ESS Imaging beamline without choppers, with cold moderator used in uni-spectral mode. The design is the same in horizontal and vertical dimension. A parobolic feeder guide focuses the beam at 6.25\,cm (aperture position, first focus point). Elliptically defocusing part followed by straight guide part before elliptically focusing into a pinhole at 50\,m (second focus point). Other instruments might place a small sample at 50\,m instead of a pinhole.}
\label{f_SkizzeImaging}
\end{figure}

The performance of different extraction systems was investigated for an example guide system chosen here to be part of an imaging beamline under consideration for ESS as shown in figure~\ref{f_SkizzeImaging}. The maximum guide width is of the order of 10\,cm. A full description of this guide can be found in \cite{ImagingGuide}. If no extraction system is used, the 60\,m long instrument is tailored to neutrons from the 12$\times$12\,cm\sq\,cold moderator and needs a narrow beam at 6.25\,m where a 1.5$\times$1.5\,cm\sq\,aperture is placed in the simulation, substituting a double-chopper planned to be used there. This focusing is achieved using a 4\,m long parabolically shaped feeder. Another focal point at 50\,m delivers a compressed beam for a 3\,cm diameter pinhole at this position, through which the 25$\times$25\,cm\sq\,sample at 10\,m distance is illuminated. The neutron beam there must have a smooth, preferably flat intensity distribution independent of wavelength. In case of a small pinhole, spatial homogeneity on the sample is equivalent to a homogenous divergence distribution in the focal point. Therefore the performance here is also a measure of the performance of similar instruments that would put a small sample at around 50\,m from the source. Since many instruments are designed this way, we will refer to the 25$\times$25\,cm\sq\,detector at 60\,m in the following instead of the imaging sample which would be placed directly in front.

The simulations used VITESS~\cite{Vitess,VitWeb} version 2.11 and a modified source matching the updated ESS cold moderator (included from VITESS~3 onwards). For the bi-spectral extraction, a 1\,cm gap between the moderators (12$\times$12\,cm\sq\,each) is assumed in the simulations. At the ESS, 1\,cm is a conservative assumption since one moderator is partly enclosed by the other such that from the view of a neutron guide, they appear to be next to each other with no gap at all. If a larger gap is needed at other neutron sources, the efficiency for cold neutrons will decrease due to the larger reflection angles and hence larger mirror coating needed. The mirror substrates are assumed to be 0.5\,mm thick silicon plates, whose attenuation and refraction properties are modeled in VITESS. For example, 1\,\Angst\,neutrons with 0\DEG\,divergence and hence an incident angle to the mirror surface of 1.175\DEG\, in case of an $m$\,=\,5 super-mirror are transmitted with 92\,\% probability. Note that the supermirror layers are assumed to be so much thinner than the substrate that attenuation in the coating is negligible. Technical problems like radiation damage are beyond the scope of this work. A uniform mirror coating of $m$\,=\,5 is used for the whole guide, including the feeder extension walls surrounding the extraction mirror(s) even if other coatings are used for the mirror(s). Reflectivity curves are modelled as linearly falling from a maximal reflectivity of 99\,\% at $m$\,=\,1 or less to a minimum reflectivity of 72\,\% at $m$\,=\,5, followed by a sharp drop to 0\,\%.  

\subsection{Performance of shortened extraction system}

\begin{figure}[tb!]
\centering
\subfigure[Spectrum at feeder entry\label{f_SpectrumEntry}]{\includegraphics[width=.45\linewidth]{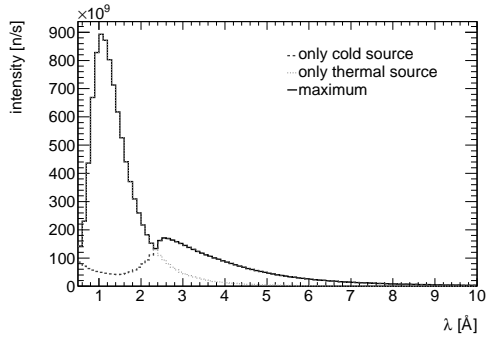}}
\subfigure[Spectrum on detector\label{f_SpectraUni_Detector}]{\includegraphics[width=.45\linewidth]{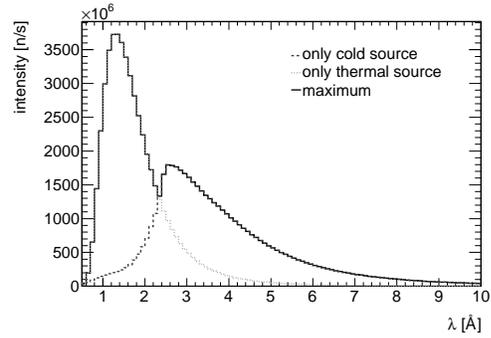}}
\caption{Spectrum at feeder entry and detector position as seen in uni-spectral mode, intensities respectively integrated over guide entry and detector area. The maximum of the cold and thermal option taken at the detector position is used to characterize the efficiency of bi-spectral extraction.}
\label{f_SpectraUni}
\end{figure}

The spectrum of neutrons entering the guide system at 2\,m and that at the detector position at 60\,m are shown in figure~\ref{f_SpectraUni} for both the thermal and cold moderator if used in ``uni-spectral mode'', i.e. using only one moderator and aligning the guide axis with the center of the respective source. From figure~\ref{f_SpectrumEntry}, the cross-over wavelength is seen to be at 2.35\,\Angst. In the following, efficiencies of various extraction systems are given at the detector position as intensity divided by the maximum of the intensity reached with the thermal or cold moderator when used in uni-spectral mode.  

\begin{figure}[tb!]
\centering
\subfigure[1 long vs. 4 short mirrors, $m$\,=\,5 \label{f_PerformanceNmirr}]{\includegraphics[width=.48\linewidth]{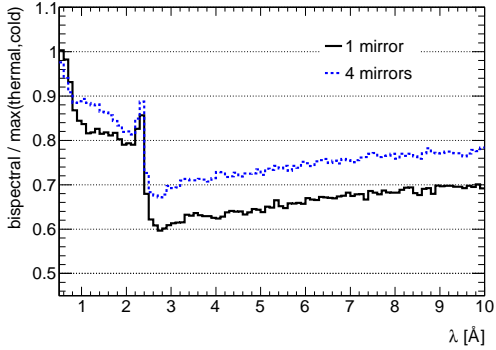}}
\subfigure[Shape of guide extension surrounding extraction mirrors (4 mirrors, $m$\,=\,5). Wall shape (a) gives only marginally different performance from the arrangement without guide walls, therefore lines lie on top of each other for most wavelengths. \label{f_PerformanceWalls}]{\includegraphics[width=.48\linewidth]{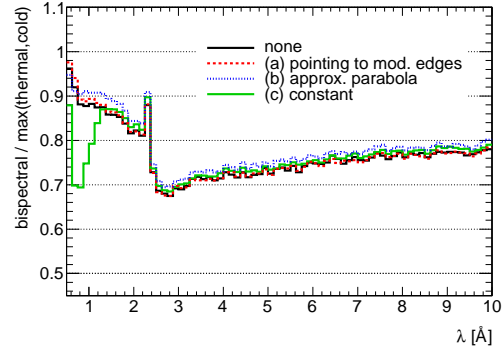}}
\subfigure[Different mirror coatings and critical angles for 2.35\,\Angst \label{f_PerformanceCoating}]{\includegraphics[width=.48\linewidth]{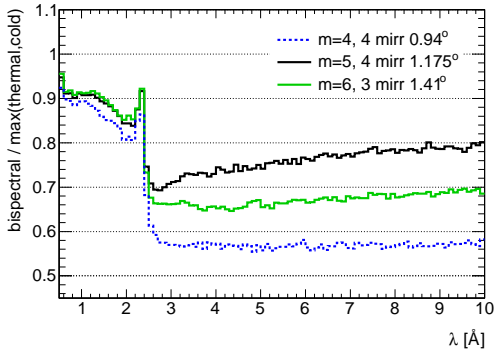}}
\subfigure[Different mirror lengths, $m$\,=\,5 \label{f_PerformanceMirrLength}]{\includegraphics[width=.48\linewidth]{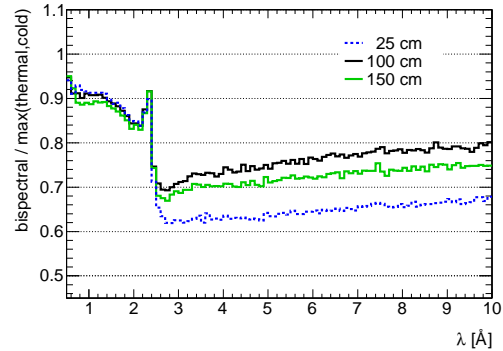}}
\caption{Efficiency of different extraction systems, calculated as the intensity on the detector divided by the maximal intensity obtainable in uni-spectral thermal or cold mode. The cross-over wavelength is 2.35\,\Angst.}
\label{f_Performance1}
\end{figure}

Figure~\ref{f_Performance1} shows this efficiency as a function of wavelength for different extraction systems. Some general features are seen in all mirror systems: The efficiency for thermal neutrons is mostly between 80\,\% and 100\,\%, but drops rapidly after a small peak at the cross-over wavelength to a minimum of 60\,\% to 70\,\% before rising again with increasing neutron wavelength up to about 80\,\% at 10\,\Angst. The efficiency for cold neutrons increases with neutron wavelength since larger reflection angles are possible, so neutrons emerging closer to the far edge of the cold moderator can be reflected into the guide system as well. Furthermore, the reflectivity for neutrons with a certain divergence increases with the neutron wavelength. The peak at the cross-over wavelength $\lambda_c$\,=\,2.35\,\Angst\,is caused by the minimum seen at this point in the reference spectrum in figure~\ref{f_SpectraUni_Detector}, with respect to which the efficiency of the smooth spectrum obtained with an extraction system is calculated. The maximal uni-spectral flux at the cross-over wavelength is exactly half the combined flux of both moderators. If the efficiency is taken with respect to the sum instead of the maximum of both moderators, no such artificial peak is observed; however most instruments must choose between bi-spectral beam extraction or a uni-spectral thermal or cold moderator. Furthermore, the extraction mirror is comparable to a switch, selecting neutrons with a certain wavelength and divergence to be taken from one moderator or the other. Therefore, displaying the efficiency with respect to the maximum is the more sensible approach. 

Figure~\ref{f_PerformanceNmirr} compares the bi-spectral extraction using just one long mirror with that of the shortened system with four 1\,m long mirrors, all using an $m$\,=\,5 coating. In both cases, the feeder is cut at the end point of the mirror(s), and its first part is replaced by straight guide walls that point to the edges of the moderators. As expected, the shortened extraction system shows better performance, yielding nearly 10\,\% more cold neutrons and a few percent more thermal neutrons. Only for wavelengths less than about 0.8\,\Angst\,is the one mirror solution slightly better. 

\begin{figure}[tb!]
\centering
\includegraphics[width=.6\linewidth]{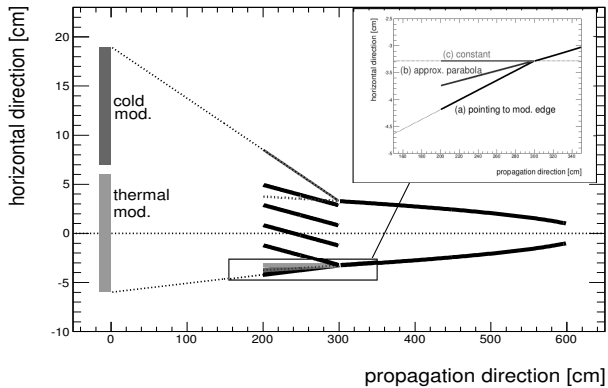}
\caption{Top view of extraction system with four straight mirrors: possible shapes of surrounding guide walls, illustrated for the horizontal wall pointing towards the edge of the thermal moderator in case (a). Extraction mirrors are not shown in zoom region. Only one vertical wall (as illustrated) and both horizontal walls (not shown) are varied.}
\label{f_SkizzeWalls}
\end{figure}

As a starting point, all guide walls surrounding the extraction system had been pointing towards the edges of the moderators in order to avoid blocking any neutrons that might otherwise have entered the guide. This principle will be kept for the vertical wall facing the cold moderator. For the other three guide extension walls, as exemplarily illustrated for the vertical wall towards the edge of the thermal moderator in figure~\ref{f_SkizzeWalls}, alternatives of constant walls parallel to the guide axis ((c) in fig.~\ref{f_SkizzeWalls}) and of walls approximating the parabolic feeder shape as used without extraction system ((b) in fig.~\ref{f_SkizzeWalls}) are compared. The performance difference can be seen in figure~\ref{f_PerformanceWalls}; the shape of the original parabolic feeder, approximated by straight guide walls through the beginning and endpoint, is seen to yield the most neutrons for all wavelengths apart from very short ones ($<$\,0.8\,\Angst). A constant guide with cross section according to the remaining feeder's entry performs second best for neutrons with wavelengths larger than 1.4\,\Angst, but is significantly worse for smaller wavelengths. Therefore in the following, a straight approximation to the parabola is used. 
Note that the 1\,m long mirror system has been used here; the performance differences are larger for longer mirrors. For the single 2.34\,m long extraction system, approximating the original feeder performs worse than a system where guide walls point towards the moderator's edges for neutrons with wavelengths shorter than 1.4\,\Angst, but is best for neutrons with $\lambda >$\,1.4\,\Angst. The constant option is by far the worst, which is not surprising since the focusing feeder's cross-section at the end of the long mirror has already decreased to about 2/3 of its value at 2\,m. 

As figure~\ref{f_PerformanceCoating} shows, the extraction efficiency with the short mirror system is best if a coating of $m$\,=\,5 is used. The number of mirrors varies between 3 and 4 here, due to the different corresponding angles leading to a different distance between mirrors depending on the mirror coating.

The mirror length of 1\,m is an arbitrary choice. The shorter the mirrors are, the more of the original guide can be used, but the smaller the extraction area illuminated by the cold moderator is and hence the steeper the incident angle of cold neutrons. This trade-off could not be resolved analytically, and so a shorter (25\,cm) and a longer (150\,cm) mirror system were simulated and their performance compared in figure~\ref{f_PerformanceMirrLength}. The initial choice of 1\,m long mirrors is the best of these three options.

\subsection{(De)focusing mirror options}
\begin{figure}[tb!]
\centering
\subfigure[Straight mirrors, decreasing incident angle.\label{f_SkizzenMirrorVariationsA}]{\includegraphics[width=.32\linewidth]{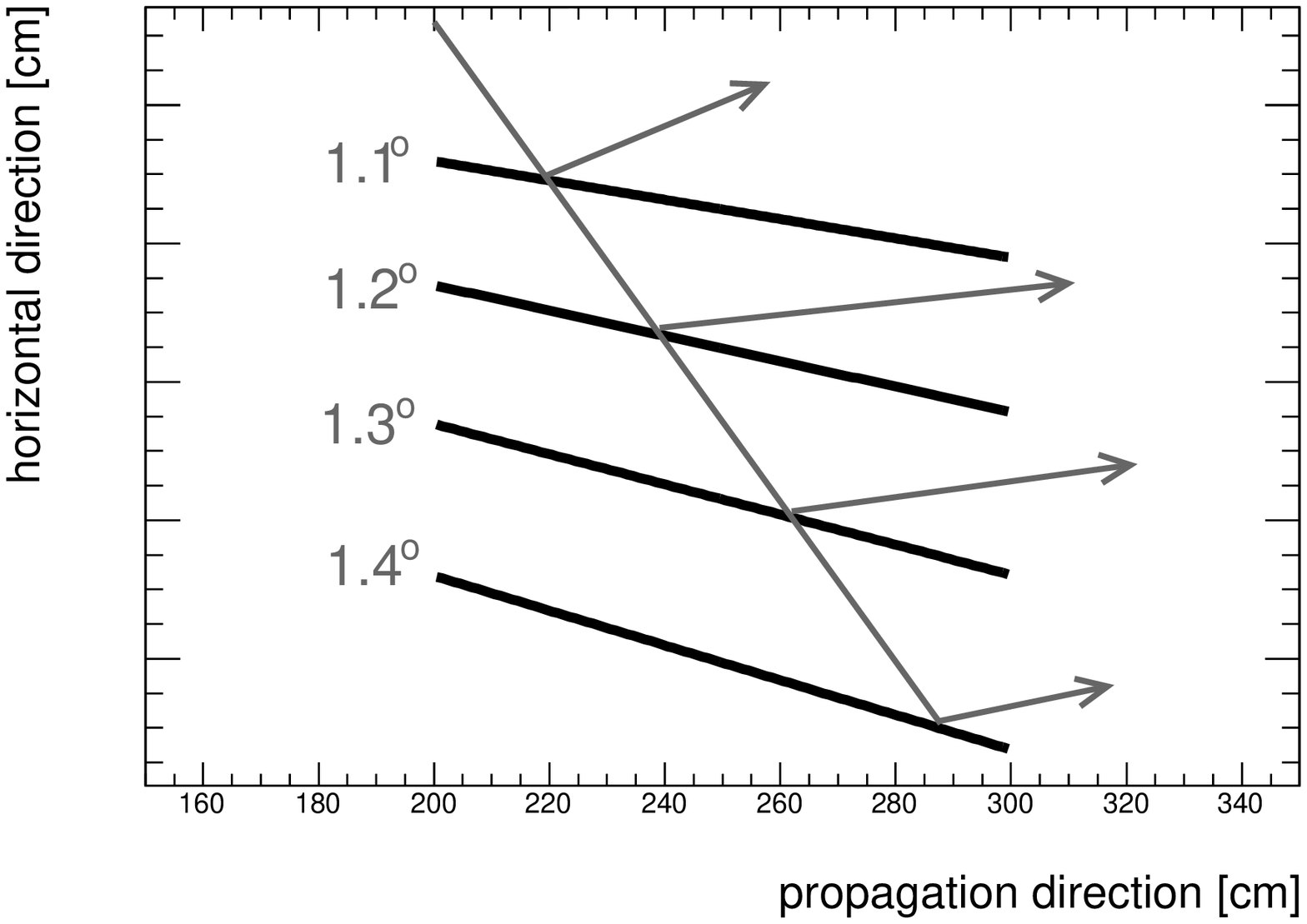}}
\subfigure[Straight mirrors, focusing.\label{f_SkizzenMirrorVariationsB}]{\includegraphics[width=.32\linewidth]{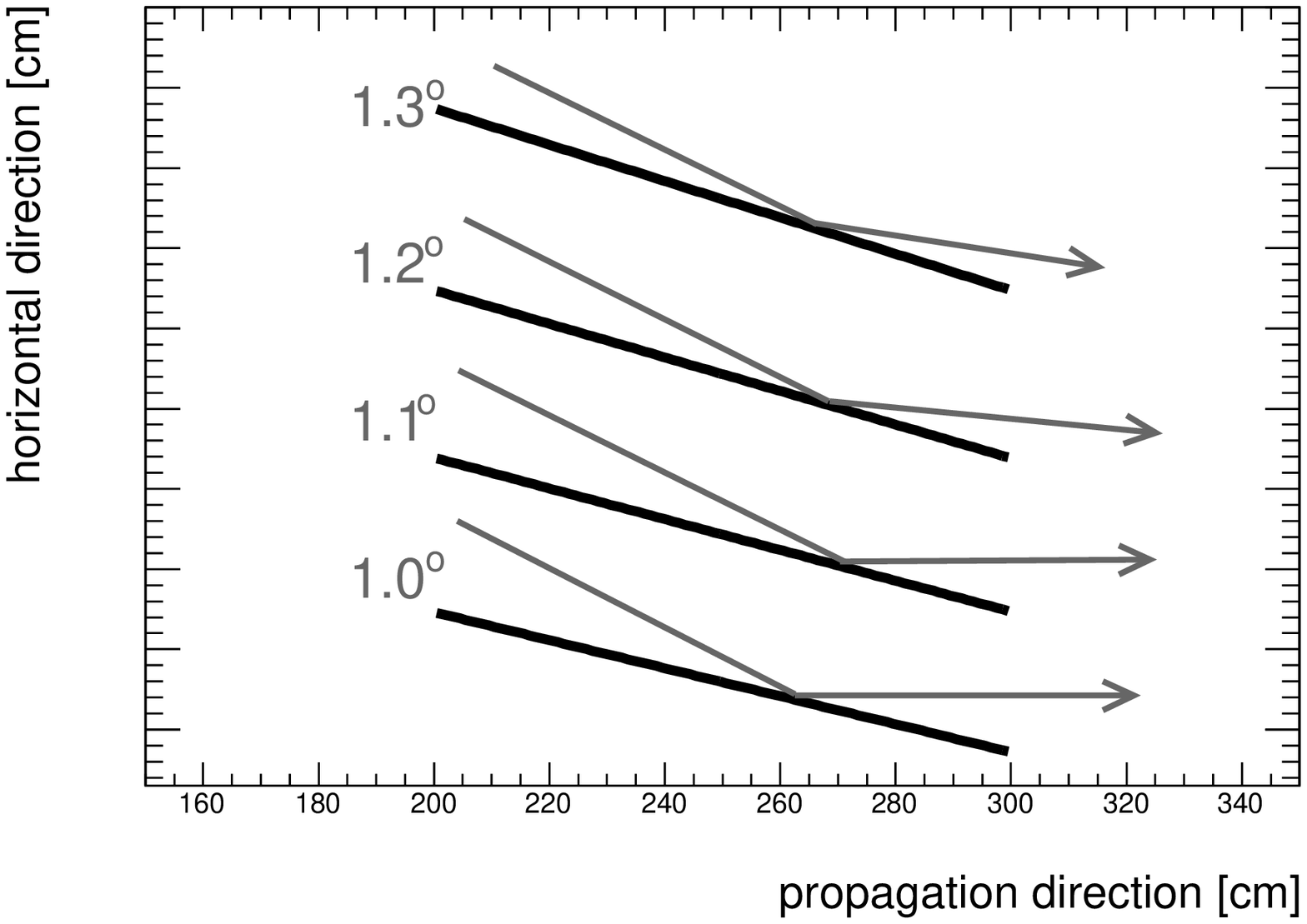}}
\subfigure[Bent mirrors, focusing. \label{f_SkizzenMirrorVariationsC}]{\includegraphics[width=.32\linewidth]{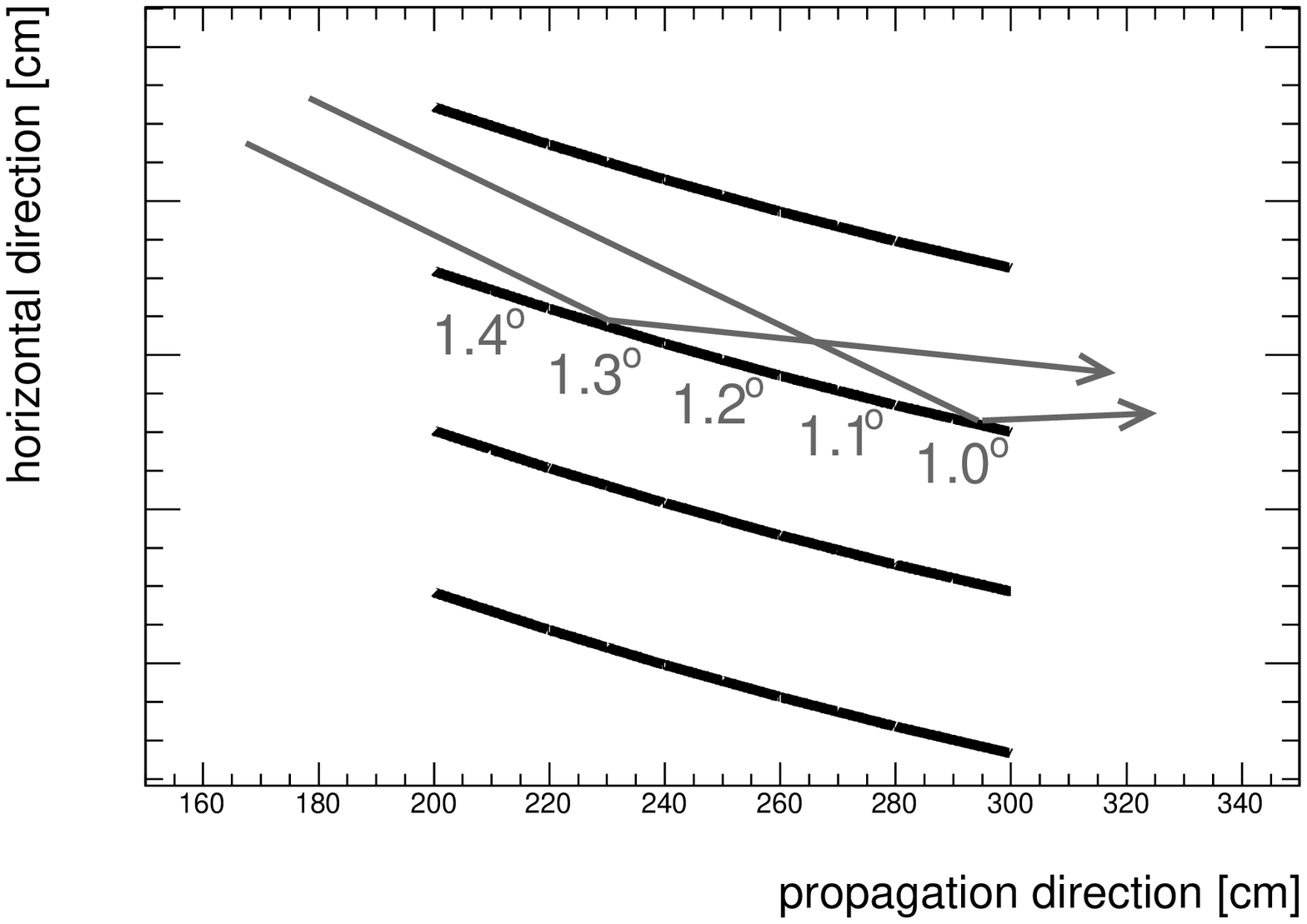}}
\caption{Schematic drawing of different extraction systems with variable mirror angles, top view.}
\label{f_SkizzenMirrorVariations}
\end{figure}

An elliptical mirror with the same focal point as the feeder can be considered as a sensible alternative for one long mirror. While this approach is not feasible in the shortened extraction system, other kind of slightly focusing or defocusing options can be investigated:

When the extraction system consists of several mirrors, the mirror inclination can be varied such that the critical angle is reached for different wavelengths, and at the same time the critical angle for the cross-over wavelength is reached for different divergences on the different mirrors. The arrangement drawn in fig.~\ref{f_SkizzenMirrorVariationsA} is designed such that the mirror furthest away from the cold moderator has the highest inclination, so the incident angle of a neutron coming from the cold moderator is getting smaller the more mirrors are passed by the neutron. The opposite design with increasing incident angles is expected to reflect less neutrons but has focusing properties instead, so more neutrons might be transported to the detector. This mirror arrangement is illustrated in figure~\ref{f_SkizzenMirrorVariationsB}. A third design in which each mirror is composed of five parts each with a slightly different angle such that the mirrors are effectively bent is shown in figure~\ref{f_SkizzenMirrorVariationsC}. 

\begin{figure}[tb!]
\centering
\subfigure[Efficiency for different (de)focusing designs, compare figure~\ref{f_SkizzenMirrorVariations}. \label{f_Eff_Focusing}]{\includegraphics[width=.48\linewidth]{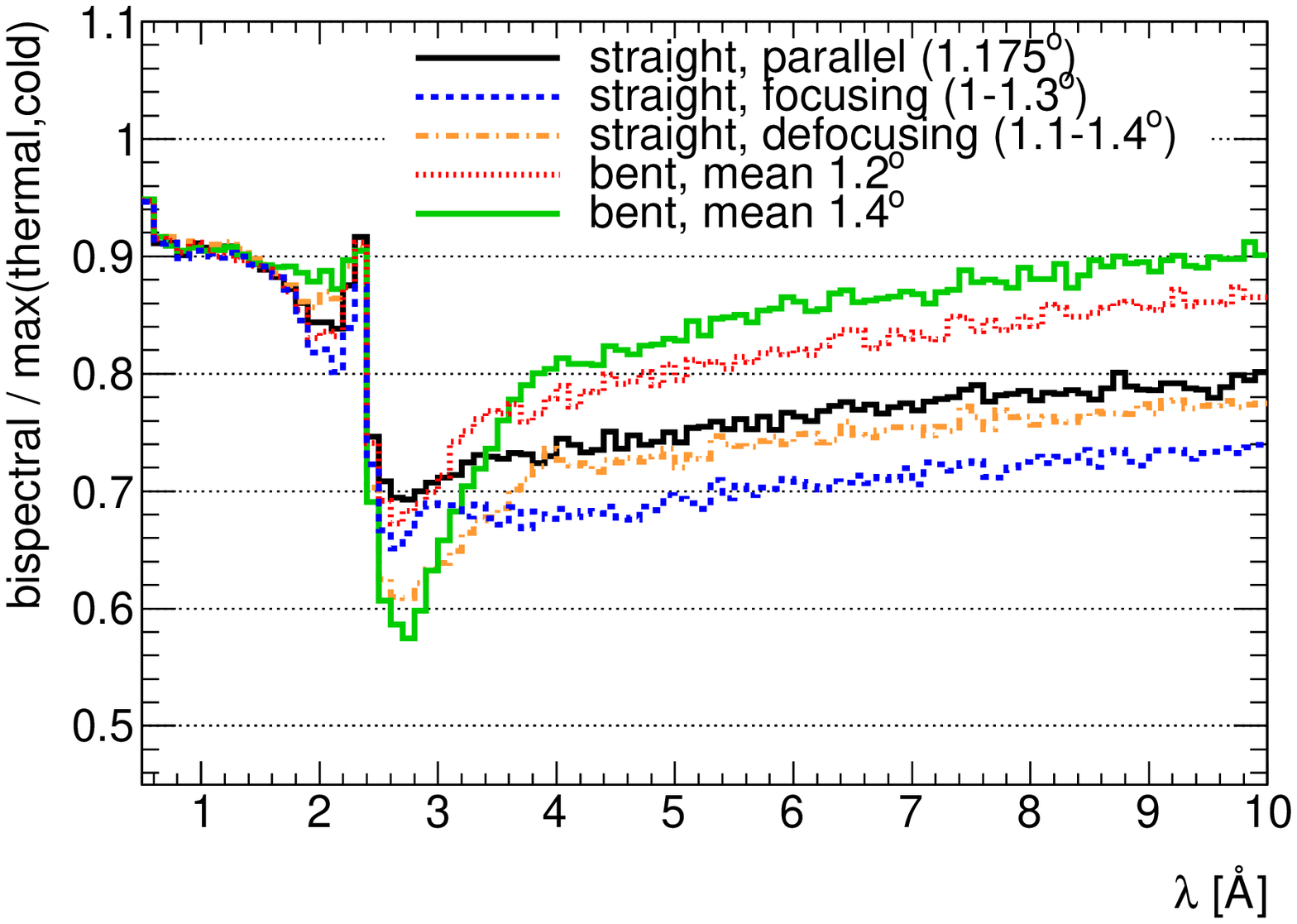}}
\subfigure[Horizontal phase space at detector position for bent mirrors with 1.2\DEG\,mean angle.\label{f_Y_Focusing}]{\includegraphics[width=.48\linewidth]{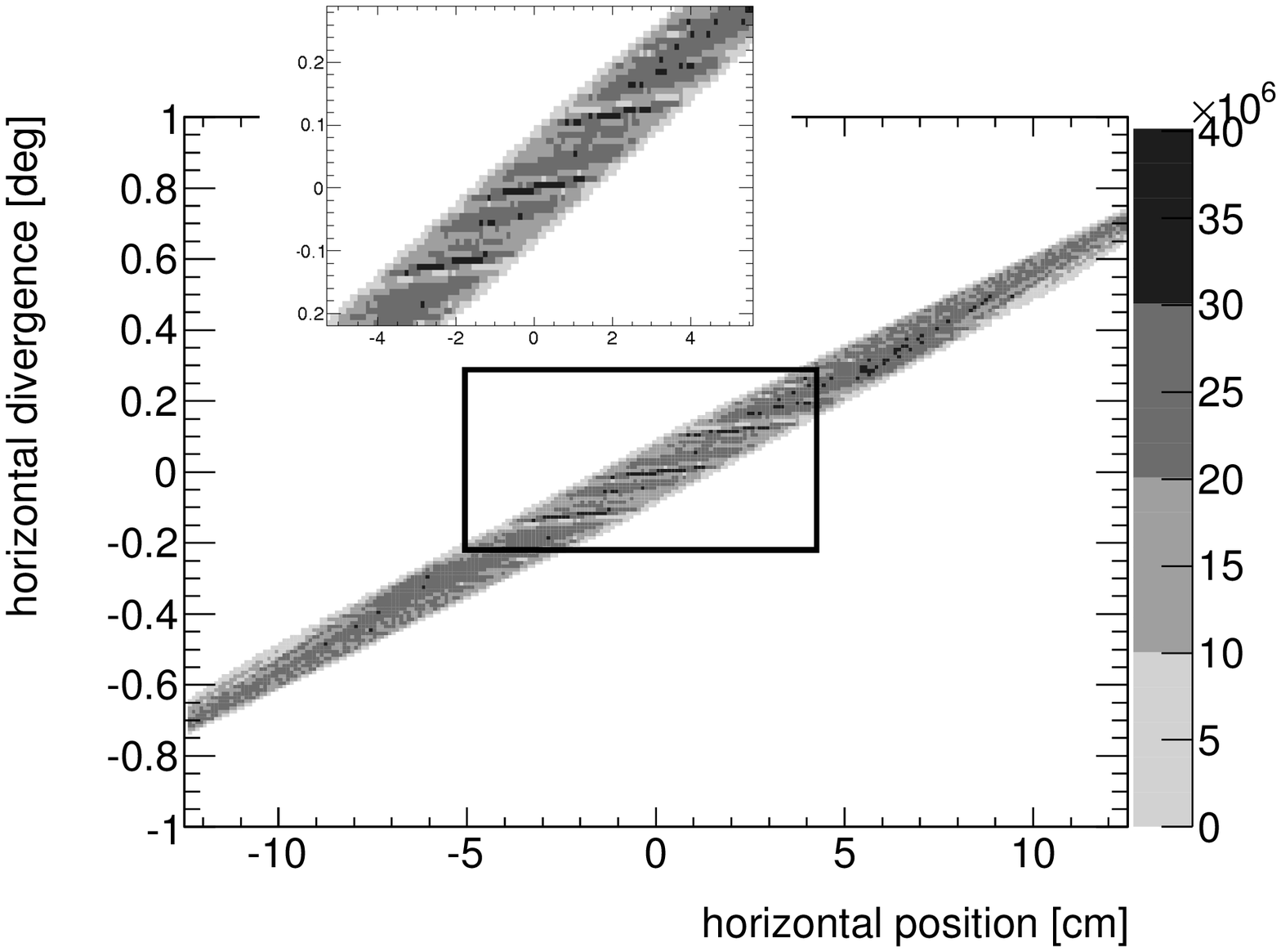}}
\caption{Efficiency at the detector position for (de)focusing mirror designs using $m$\,=\,5 and impact on beam homogeneity for the system chosen for the further study. The angular difference between single mirrors in the straight mirror options as well as between the five mirror parts in the bent options is 0.1\DEG. See text for further details.}
\label{f_PerformanceFocusing}
\end{figure}

The performance of these systems is shown figure~\ref{f_PerformanceFocusing}. In addition to the obtained spectrum (normalized to the maximum obtained without extraction system), the horizontal phase space of the bent mirrors with a mean angle of 1.2\DEG\,is shown. Straight mirrors tilted in a (de)focusing way deliver the highest efficiency if the difference in $\alpha$ is small, therefore the performance is only shown for angular differences of 0.1\DEG\,and mean angles around 1.2\DEG. The parallel option is seen to be better throughout the whole wavelength band with a small exception around 2\,\Angst\,where the defocusing option (as in fig.~\ref{f_SkizzenMirrorVariationsA}) is slightly better. This option performs better for cold neutrons than the focusing option, but shows a more pronounced efficiency drop between 2.5\,\Angst\,and 3.5\,\Angst. 
Compared to straight parallel mirrors, slightly bent mirrors give significantly higher efficiencies for cold neutrons but a lower efficiency between 2.5\,\Angst\,and 3.5\,\Angst. This behavior is more pronounced for larger angles. The two examples shown in figure~\ref{f_PerformanceFocusing} have varying angles of 1.4\DEG-1.0\DEG\,and 1.6\DEG-1.2\DEG; a stronger bending covering a larger angular range has also been simulated but found to deliver slightly less thermal and significantly less cold neutrons.

Neutrons with wavelengths around 3\,\Angst\,are important for many instruments; therefore we continue with the bent mirror system with a mean angle of 1.2\DEG, although this option delivers about 4\,\% less cold neutrons than bent mirrors with a mean angle of 1.4\DEG.

 A small spatial-angular correlation is observed in the central part of the detector with this system, the main part of which, the three distinct lines in figure~\ref{f_Y_Focusing}, is however not introduced by the bi-spectral extraction but also present in the uni-spectral approach. The bi-spectral extraction system only adds some fine-structure in between those lines, which is also present with straight parallel mirrors, leading to five narrow central peaks in the divergence profile between $\pm$0.15\DEG\,where, without the extraction system, only two peaks are observed. This effect is small and should not pose a major problem for most applications. When integrated over the divergence, no spatial inhomogeneity is left, and spectral differences are also observed to be small. 

\subsection{Choice of guide axis}
\begin{figure}[tb!]
\centering
\subfigure[Efficiency for different guide axis shifts.\label{f_Eff_Axis}]{\includegraphics[width=.48\linewidth]{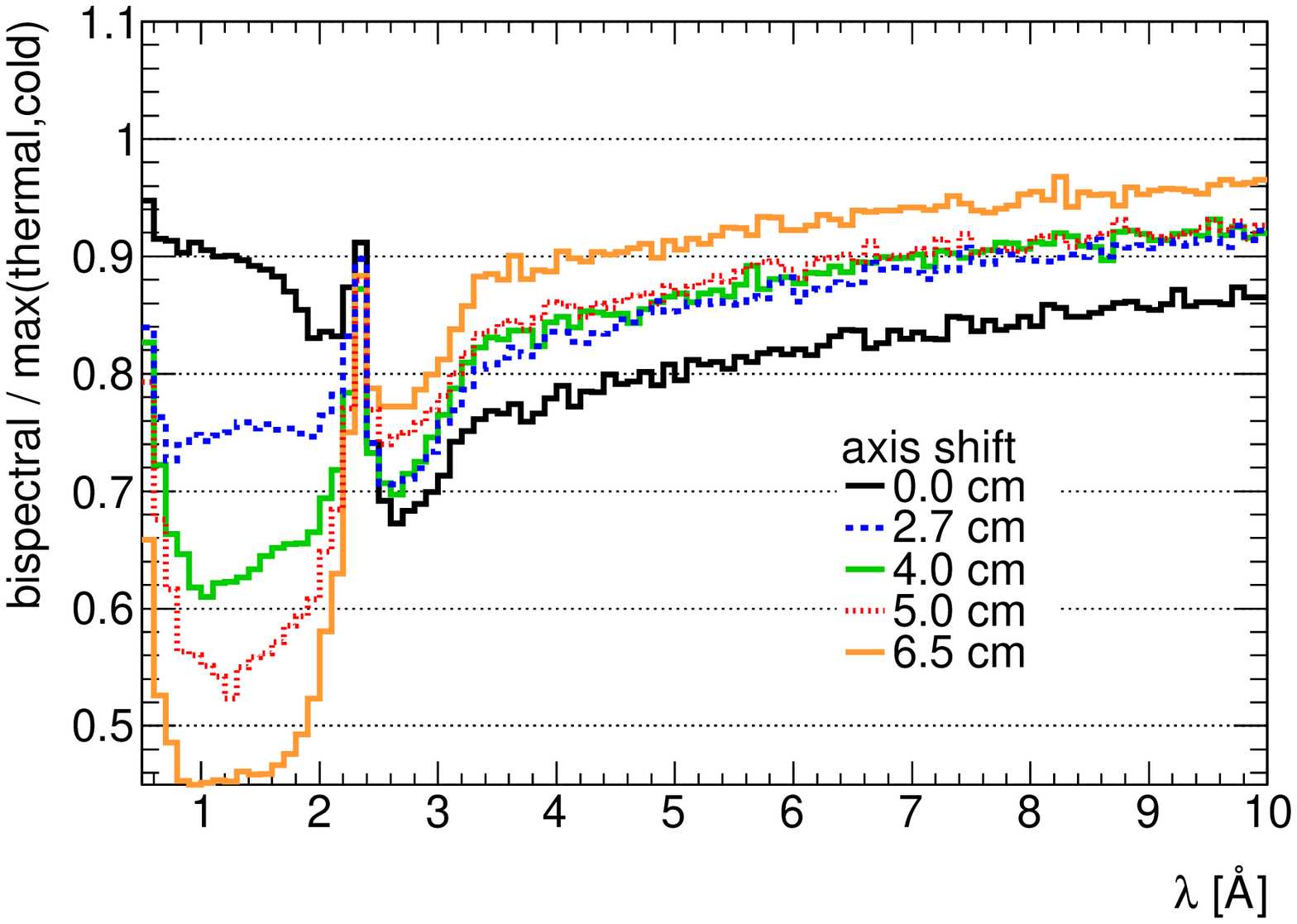}}
\subfigure[Horizontal spatial beam homogeneity for different guide axis shifts.\label{f_Y_Axis}]{\includegraphics[width=.48\linewidth]{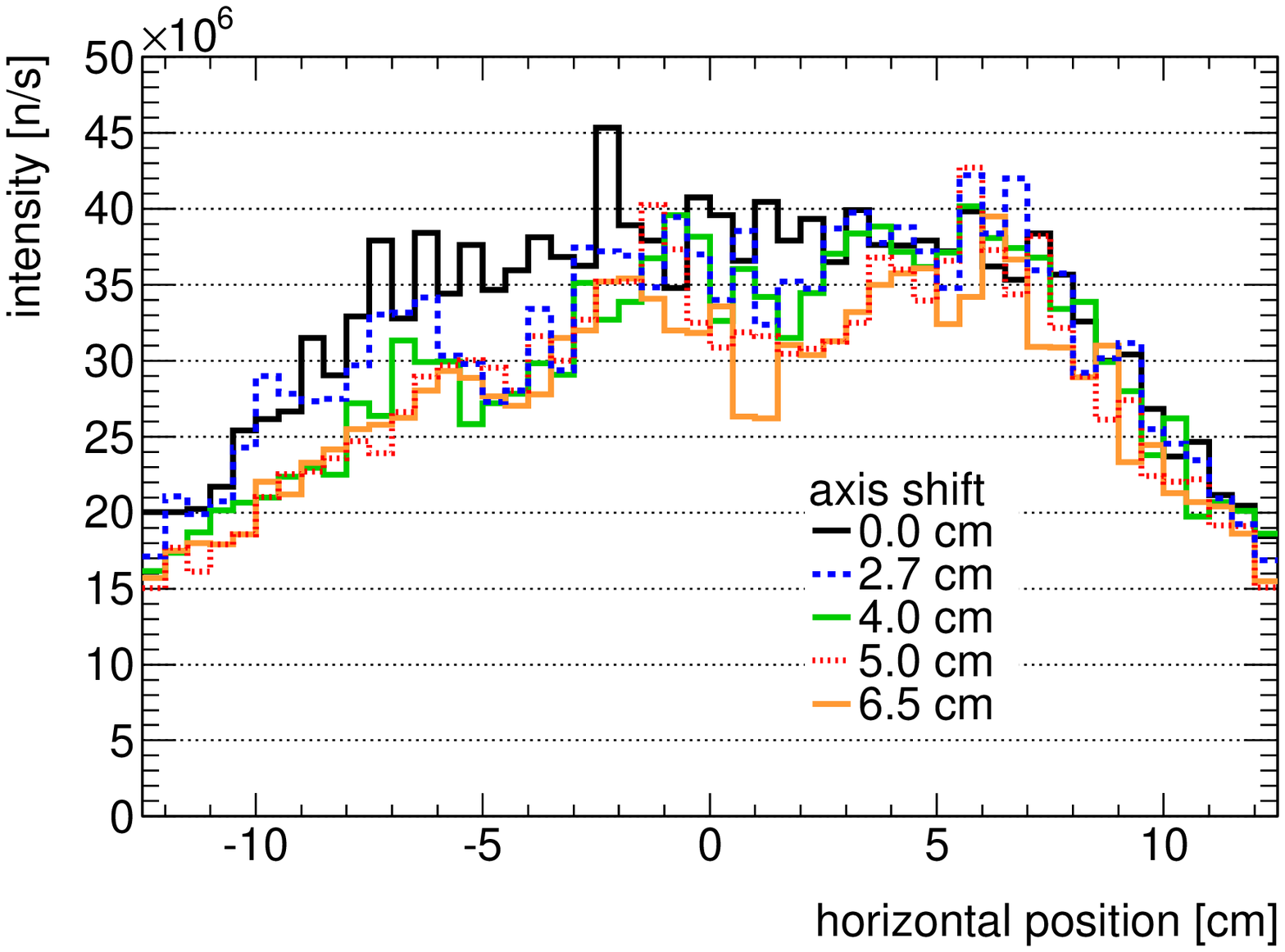}}
\caption{Efficiency and impact on beam homogeneity for different guide axes. The horizontal intensity distribution is taken in a 5\,mm central vertical strip (wavelength integrated). A trade-off between cold neutron intensity and thermal neutron intensity as well as beam homogeneity is visible.}
\label{f_PerformanceAxis}
\end{figure}

All extraction systems discussed so far have an axis (defined by the following guide system) which passes through the center of the thermal moderator. In order to extract more cold neutrons in such a way that they can be transported to the detector, the guide system and bi-spectral extraction can be shifted horizontally relative to the source towards the cold moderator. One sensible choice is to align the edge of the feeder with the inward edge of the thermal moderator, but further shifts might also be useful options depending on the relative importance of certain neutron wavelengths. Figure~\ref{f_PerformanceAxis} shows the spectra for several options of guide axis shifts as well as the spatial homogeneity. A shift of 2.72\,cm corresponds to aligning the feeder edge with the thermal moderator edge, a shift of 6.5\,cm aligns the guide system axis with the center between the two moderators. All mirror parts that would have been shifted further than the inward edge of the cold moderator have been removed, so for example in the 6.5\,cm shift option, only two of the four bent mirrors remain. 

\begin{figure}[p]
\centering
\subfigure[Position on detector, shift 0\,cm \label{f_CompareAxesA}]{\includegraphics[width=.48\linewidth]{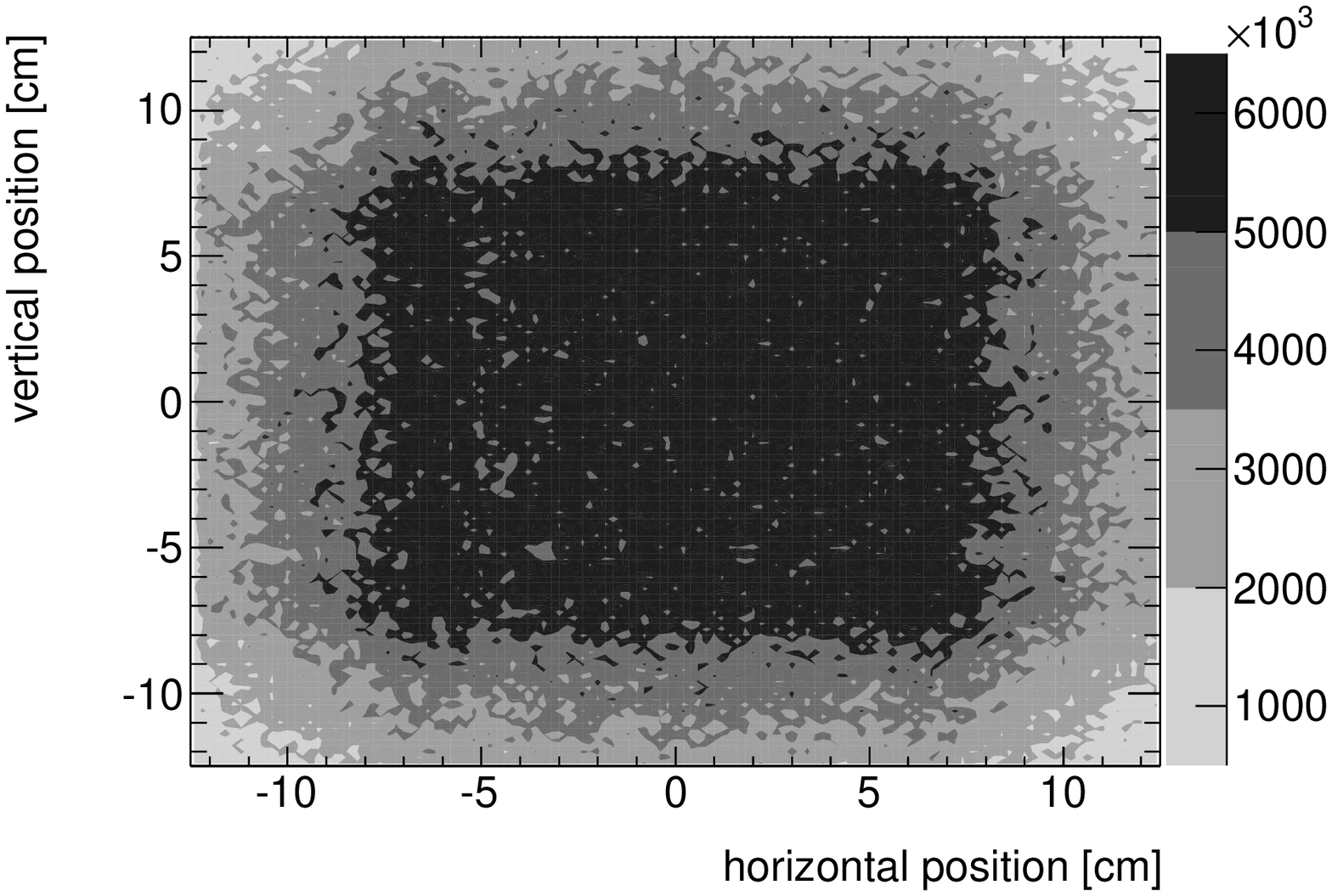}}
\subfigure[Wavelength vs. horizontal position, shift 0\,cm\label{f_CompareAxesB}]{\includegraphics[width=.48\linewidth]{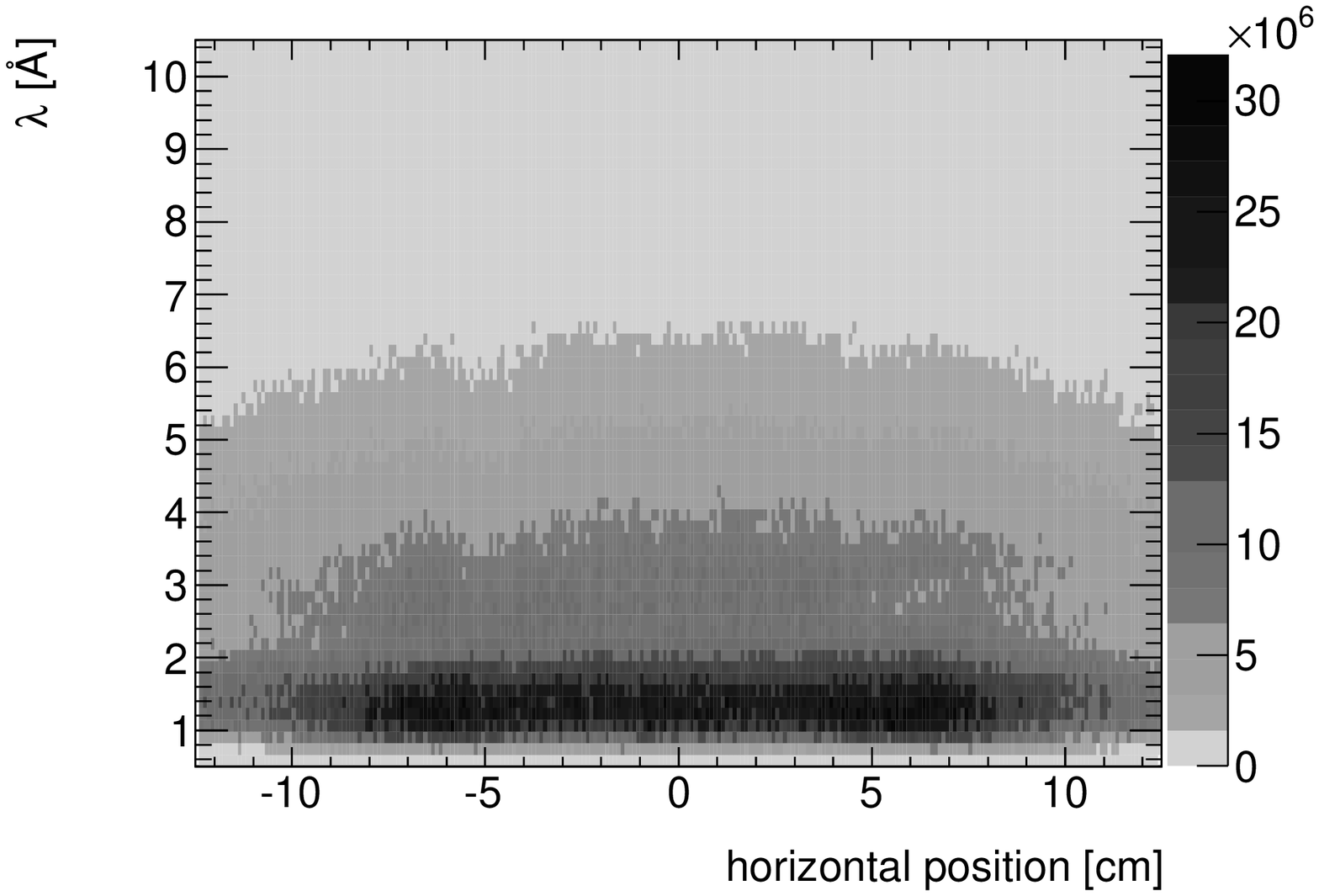}}
\subfigure[Position on detector, shift 6.5\,cm\label{f_CompareAxesC}]{\includegraphics[width=.48\linewidth]{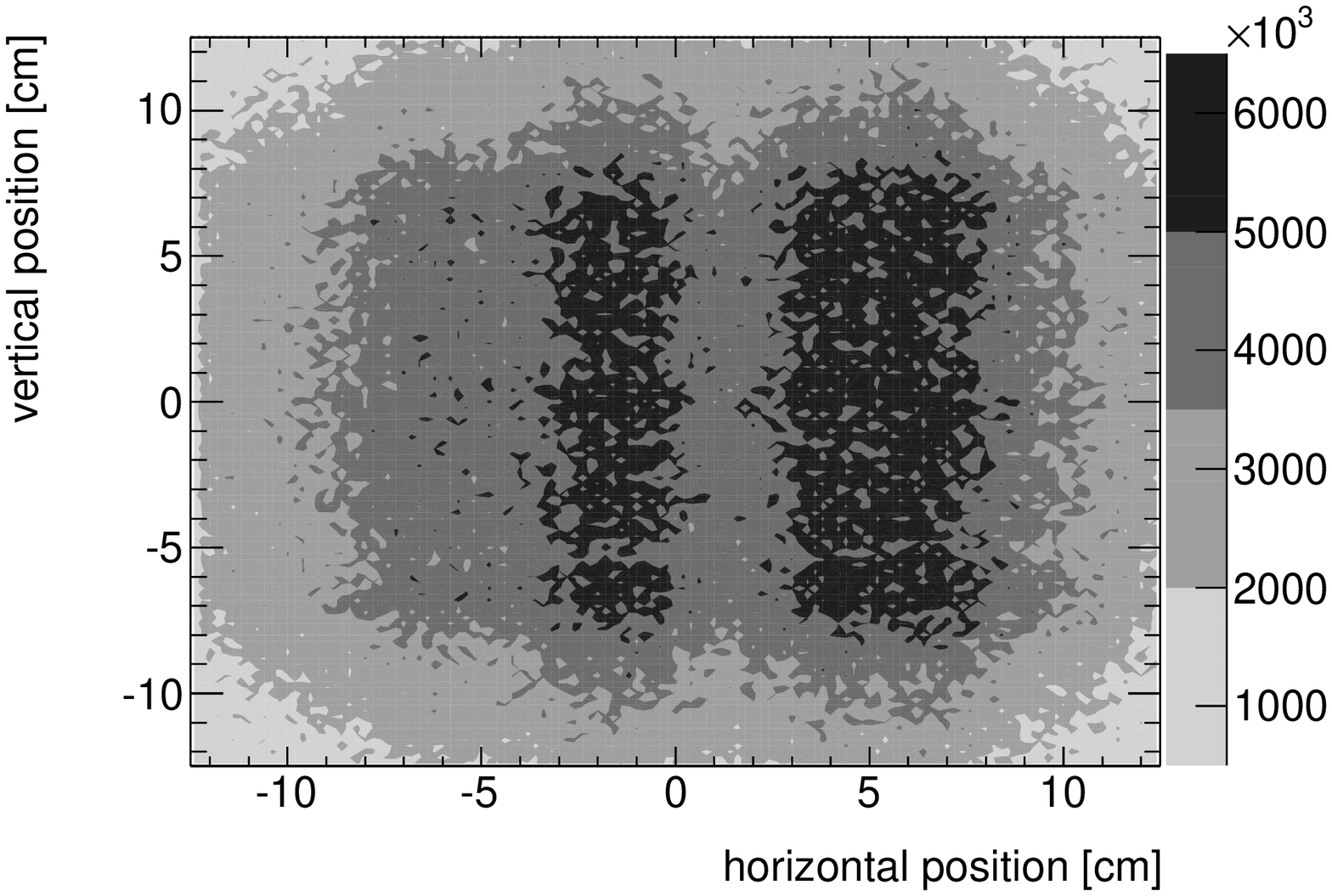}}
\subfigure[Wavelength vs. horizontal position, shift 6.5\,cm\label{f_CompareAxesD}]{\includegraphics[width=.48\linewidth]{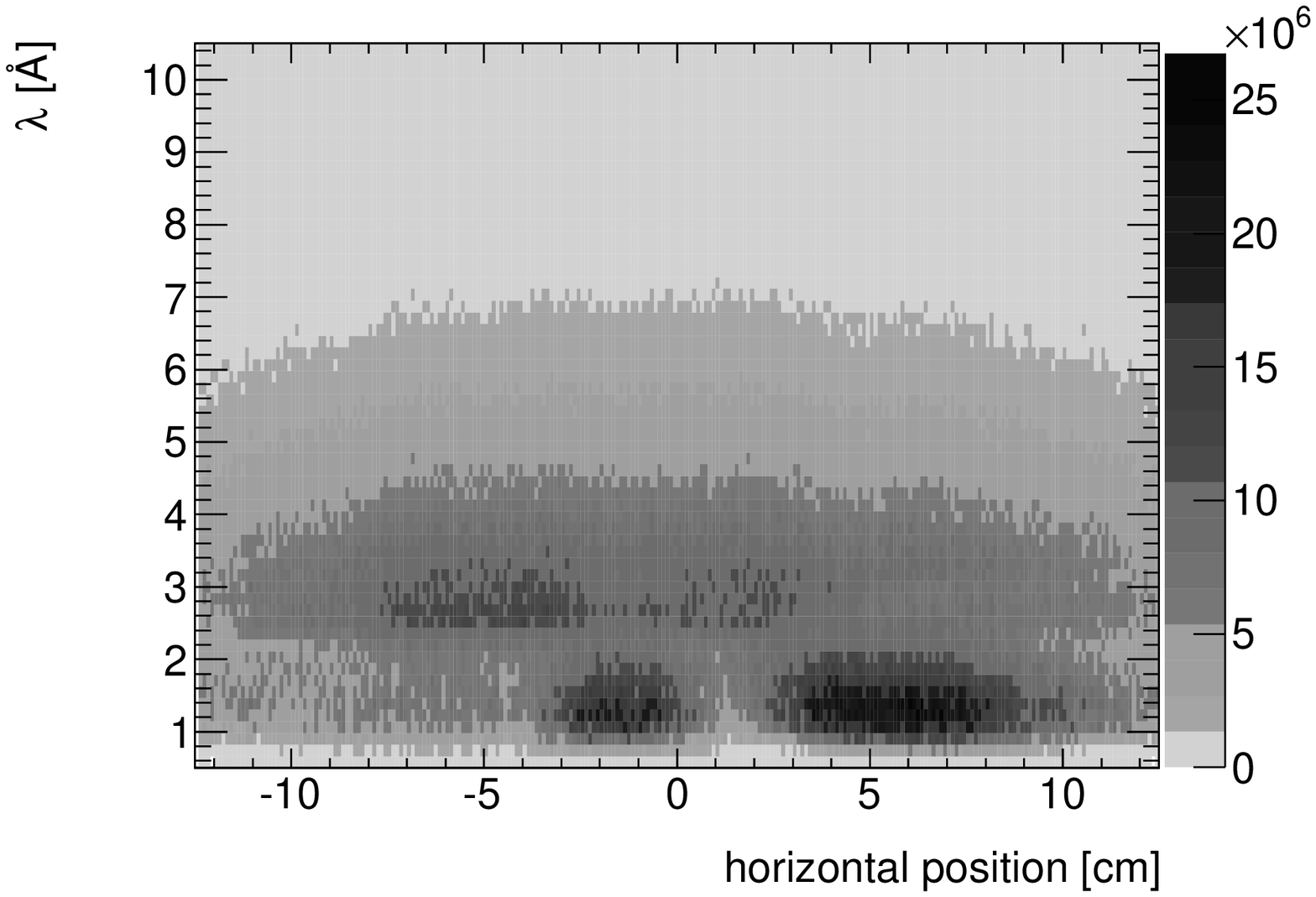}}
\caption{Impact of a guide axis shift on the beam homogeneity: Intensity [n/s] on the detector depending on vertical position (left) and wavelength (right) vs. horizontal position, for axis shifts of 0\,cm (upper row) and 6.5\,cm (lower row). Bent mirrors with a mean angle of 1.2\DEG\,are used, 4 mirrors in case of 0\,cm shift and 2 mirrors in case of 6.5\,cm shift. Severe spatial as well as spectral inhomogeneities are observed for the shifted guide axis.}
\label{f_CompareAxes}
\end{figure}

The extraction of cold neutrons can be significantly improved by such an axis shift up to a maximum of 95\,\% at 10\,\Angst, but the fractional loss of thermal neutrons is larger than the fractional gain in cold neutrons. Furthermore, the further the axis is shifted, the less homogeneous becomes the beam distribution at the detector. The relative importance of certain wavelength ranges as well as the requirements on spectral and spatial beam homogeneity may vary between different instruments, so no generally best option can be decided on here. Figure~\ref{f_CompareAxes} shows the two-dimensional intensity on the detector as well as the wavelength dependence on the horizontal beam position for the two extrema 0.0\,cm and 6.5\,cm shift to illustrate the extent of spatial as well as spectral inhomogeneity. Unsurprisingly, the most severe inhomogeneity in the intensity distribution on the detector is seen for thermal neutrons. Small divergence neutrons from the thermal moderator only illuminate half the guide system in the 6.5\,cm shift example, and thermal neutrons illuminating the far half of the guide have divergences too large to be transported efficiently. 

In summary, with a bi-spectral extraction system combined with a focusing feeder, up to 95\,\% of the cold neutrons available for the corresponding uni-spectral approach can be used if inhomogeneities in the obtained neutron beam are acceptable and about 45\,\% of the uni-spectral thermal neutrons are sufficient. Compared to the uni-spectral cold usage, this corresponds to an increase of thermal neutrons of a factor of 8. If a high beam homogeneity is important and/or both thermal and cold neutrons are desired in large amounts, a non-zero (or not much) shifted scheme is recommended, yielding up to 85\,\% of the long wavelength neutrons and about 80\,\%\,-\,90\,\% thermal neutrons. In all cases, the efficiency curve shows a minimum at around 2.5\,\Angst, which can be as low as 68\,\%.

Note that in this study, the remaining 3\,m of the feeder after the bi-spectral extraction system have been used unchanged from the feeder shape obtained from optimization of the uni-spectral cold operation mode. Including the bi-spectral extraction system in a new optimization for a particular wavelength band of interest might be helpful to adjust the feeder in the composite system to the needs of a particular instrument. This could improve the efficiency further, but such an additional optimization is beyond the scope of this work.

\subsection{Fourth guide wall and final adjustments}

As shown above, the highest extraction efficiency for cold as well as thermal neutrons is achieved if the outer walls of the extraction system, apart from the one facing the cold moderator, approximate the shape of the original feeder. For the fourth wall, this is not possible since such a guide wall would block all neutrons from the cold moderator. Simulations show that this fourth mirrored wall provides no benefit even if it points towards the far edge of the cold moderator or if its inclination is slightly decreased compared to that. 

\begin{figure}[tb!]
\centering
\subfigure[Reduced mirror system: dotted lines indicate non-contribution mirror parts that can be removed.\label{f_CollisionPoints}]{\includegraphics[width=.48\linewidth]{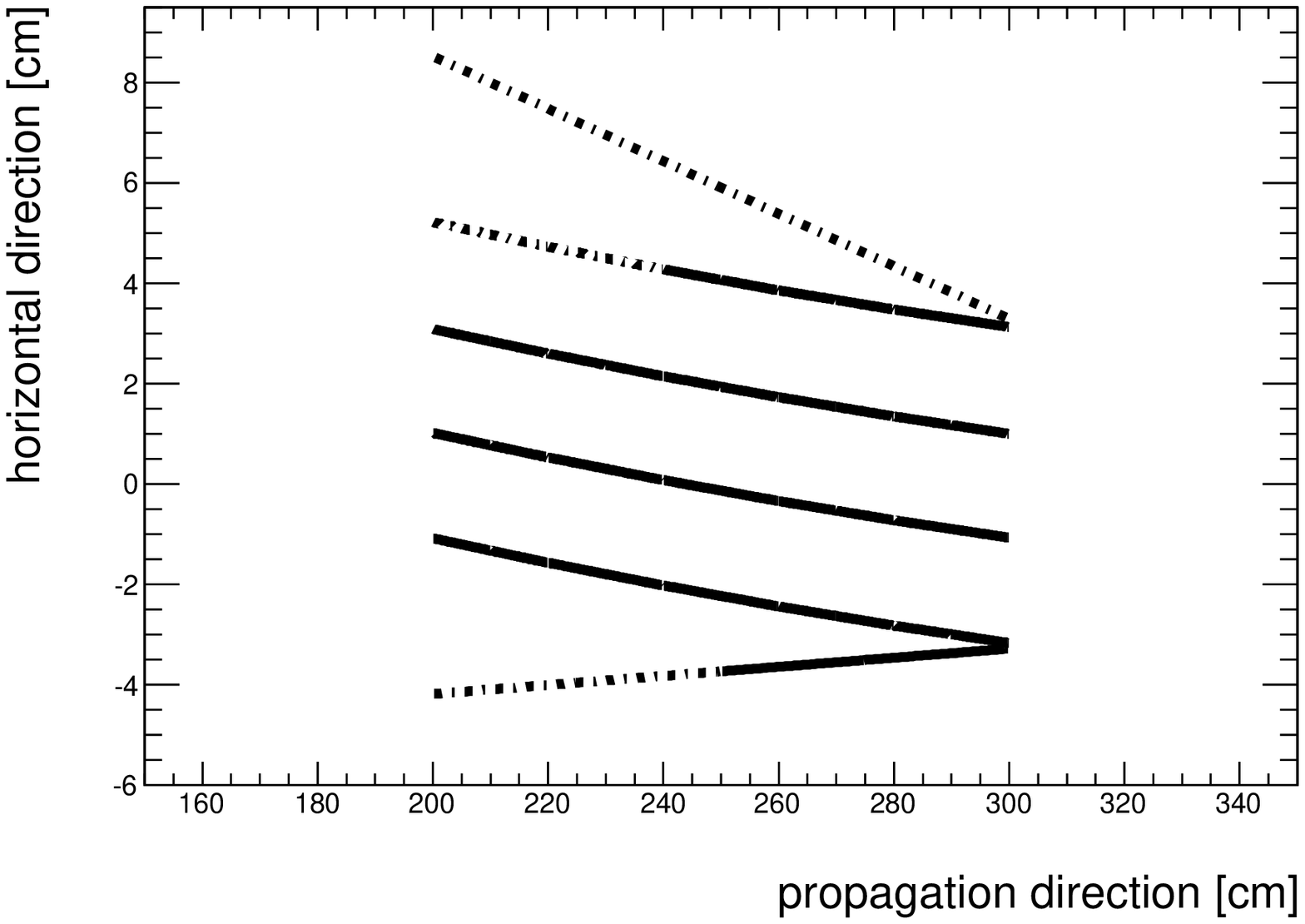}}
\subfigure[Efficiency with partly removed mirrors: no performance difference, hence curves lie on top of each other. \label{f_ReducedMirrSyst}]{\includegraphics[width=.48\linewidth]{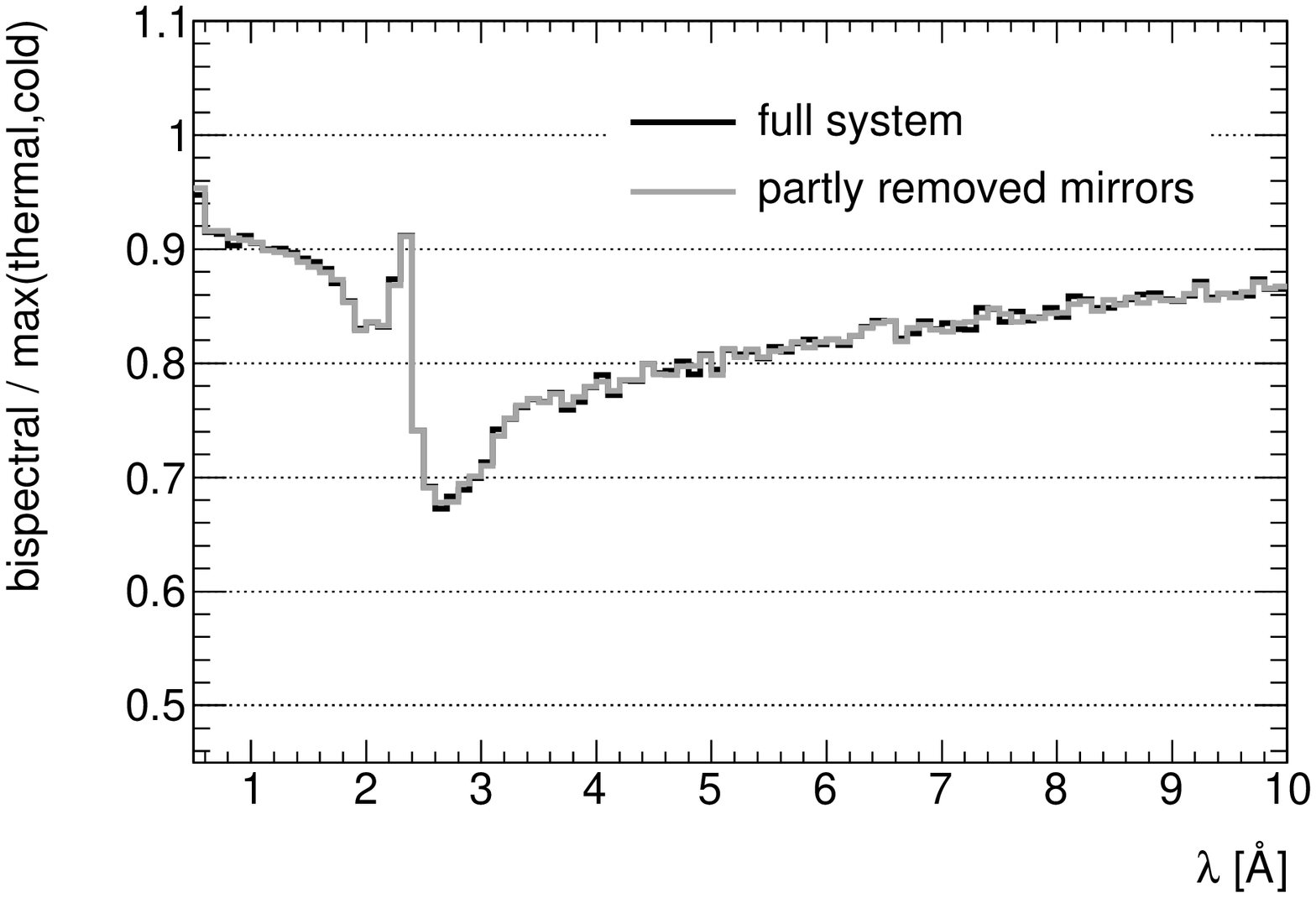}}
\caption{Non-contributing mirror parts found by traced neutrons that reach the detector (a) and efficiency when unused parts of the extraction system are removed (b).}
\label{f_MirrReduction}
\end{figure}

\begin{figure}[tb!]
\centering
\subfigure[Spectrum obtained with bi-spectral system compared to uni-spectral cold and thermal options. \label{f_SpectrumFinalA}]{\includegraphics[width=.45\linewidth]{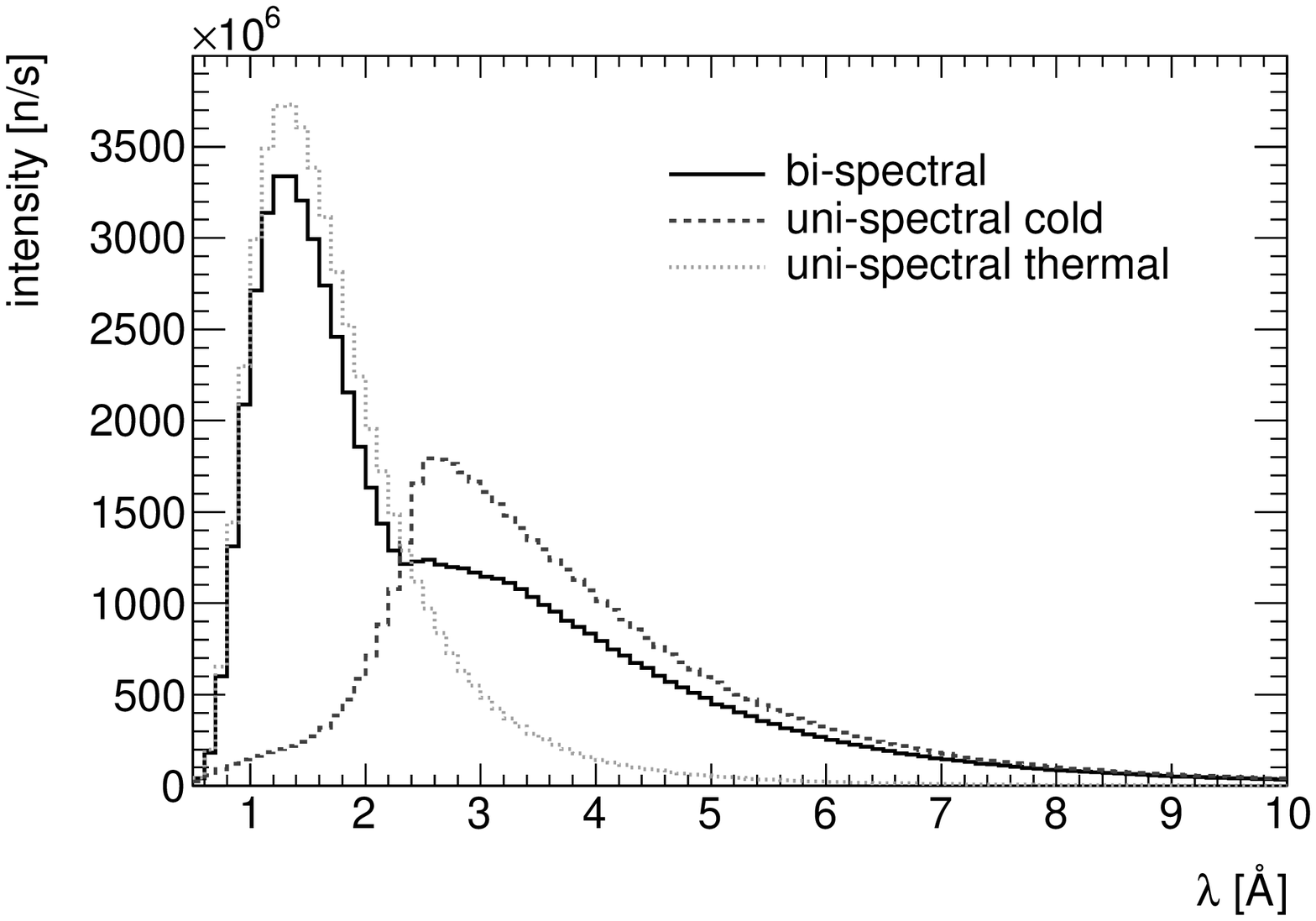}}
\subfigure[Spectrum split into contributions from cold and thermal moderator when used in bi-spectral system. \label{f_SpectrumFinalB}]{\includegraphics[width=.45\linewidth]{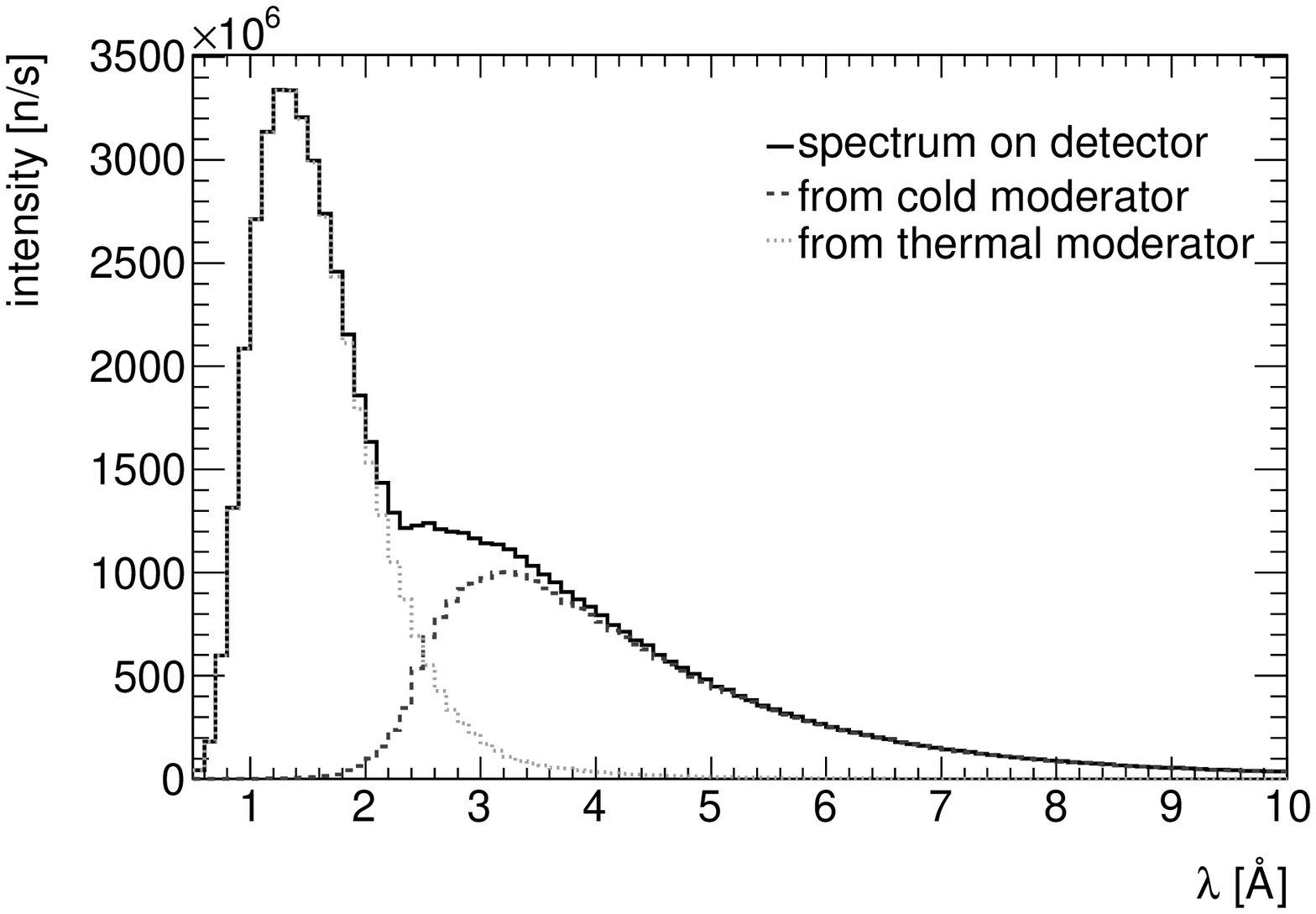}}
\caption{Spectrum at detector position with bi-spectral extraction system with 4 bent mirrors as shown in figure~\ref{f_CollisionPoints}, intensity integrated over detector area. }
\label{f_SpectrumFinal}
\end{figure}

\begin{figure}[tb!]
\centering
\includegraphics[width=.48\linewidth]{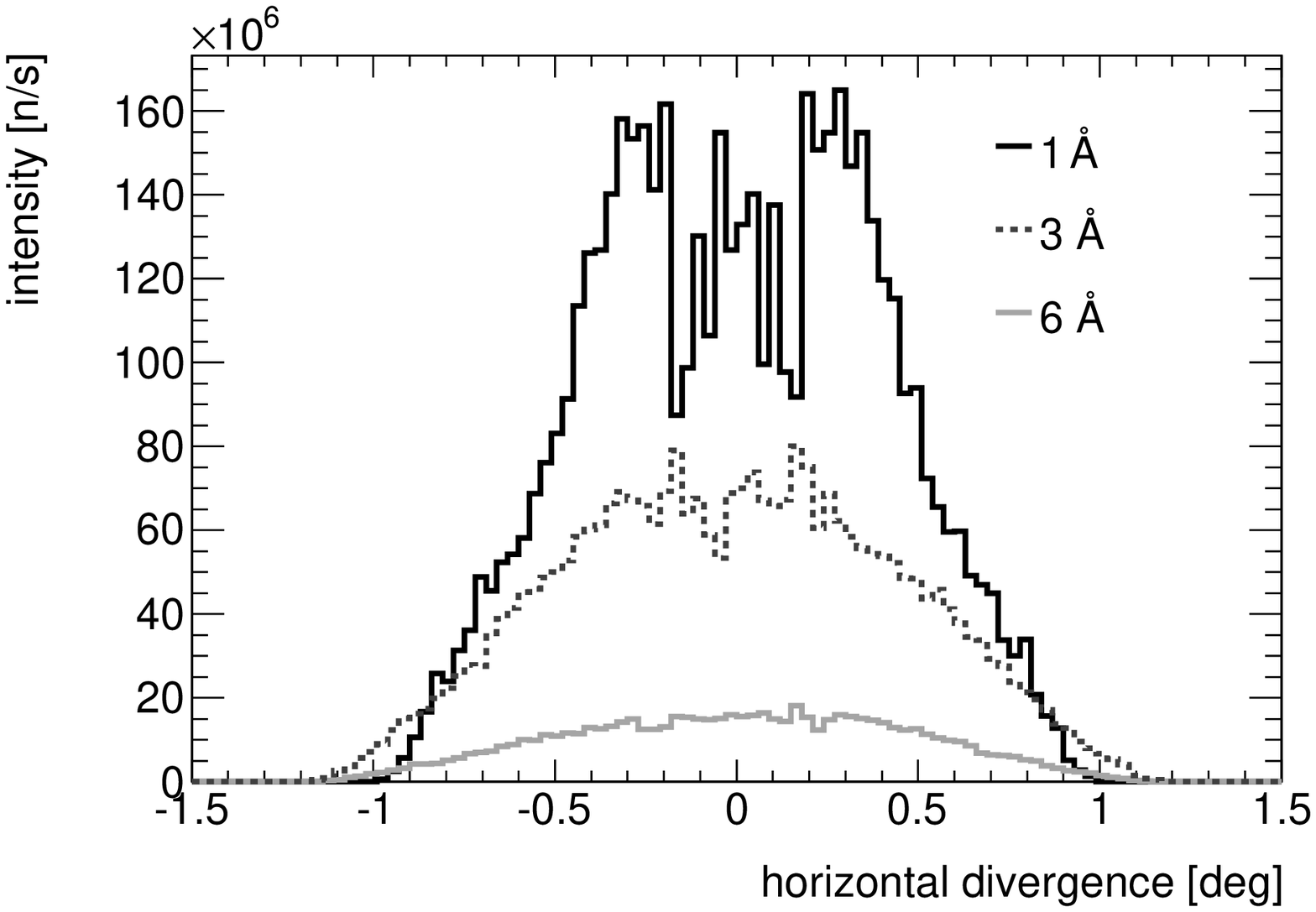}
\caption{Wavelength dependent angular distribution obtained with bi-spectral extraction system as shown in figure~\ref{f_CollisionPoints}: horizontal divergence in second focus point at pinhole position for neutrons with 1\,\Angst, 3\,\Angst\,and 6\,\Angst\,(0.2\,\Angst\,window). The observed structure for short wavelengths is expected from the elliptical focusing.}
\label{f_DivY_Focus2}
\end{figure}

In figure~\ref{f_CollisionPoints} the contributing and non-contributing parts of the mirror system are distinguished as solid and dotted lines. The non-contributing parts have been found by tracing the neutrons through the whole instrument and marking interaction points of only those neutrons that reach the detector. The guide wall pointing towards the cold moderator's edge is not contributing at all and can be removed. Similarly, the guide wall towards the thermal moderator can be cut in half, and the first 40\,cm of the mirror closest to the cold moderator can be removed as well. Figure~\ref{f_ReducedMirrSyst} confirms that indeed no efficiency loss occurs if these parts of the extraction system are removed. The spectrum obtained on the detector with this final design is shown in figure~\ref{f_SpectrumFinal} in comparison with uni-spectral options as well as with contributions from the thermal and cold moderator in the bi-spectral option marked separately. A possible wavelength dependence of the angular distribution is disproved in figure~\ref{f_DivY_Focus2}, which shows the horizontal divergence in the second focus point where a pinhole is placed in an imagine beamline and a sample could be in other instruments. The structure seen for wavelengths shorter than 3\,\Angst\,is characteristic for elliptically shaped neutron guides, as has been shown and explained in \cite{MultipleReflectionsInEllipticNeutronGuides}. No additional structure is observed. The resulting wavelength dependent spatial distribution is the same as shown in figure~\ref{f_CompareAxesB} in the bi-spectral system without axis shift, where the only difference to the final design is that non-contributing mirror parts are still included in the simulation. Similarly, the horizontal phase space at the detector position is unchanged from what is shown in figure~\ref{f_Y_Focusing}. The structure seen there is more pronounced for short wavelengths and gets washed out with longer wavelengths.

\section{Conclusions}

For many instruments, bi-spectral beam extraction can be a useful alternative to a uni-spectral beam. This work shows that for guide systems using a focusing feeder, a short extraction system consisting of several mirrors performs better than the conventional one mirror system. The ideal length for such a mirror system seems to be around 1\,m in the arrangements tested, and the best performance is obtained for a coating of $m$\,=\,5. Even for this comparably short extraction system, additional guide walls around the mirror systems are seen to affect the extraction efficiency, and are best designed to approximate an extension of the following guide shape. For most purposes, slightly bent mirrors with a mean inclination equal to the critical angle of the cross-over wavelength deliver the best performance, with a maximal cold neutron efficiency of 85\,\% at 10\,\Angst, a maximal thermal neutron efficiency of about 90\,\% for $\lambda <$\,1.5\,\Angst\,and a minimum efficiency of 68\,\% around 2.5\,\Angst. Cold neutron extraction can be increased but with a cost in neutrons with wavelengths around 2.5\,\Angst\,by changing the mean mirror angle. If the homogeneity requirement of the neutron beam is not strong, an easy method to adjust the relative extraction and transportation efficiency of cold and thermal neutrons is to vary the position of the guide system's axis. In this way, cold neutron efficiencies of up to 95\,\% can be reached.


\section{Acknowledgments}

We thank K. Lefmann, H. Jacobsen, N. Kardjilov, A. Hilger, M. Russina and K. Rolfs for fruitful discussions.

This work was funded by the German BMBF under ``Mitwirkung der Zentren der Helmholtz Gemeinschaft und der Technischen Universit{\"a}t M{\"u}nchen an der Design-Update Phase der ESS, F{\"o}rderkennzeichen 05E10CB1''.

\label{Bibliography}
\bibliographystyle{unsrtnat}
\bibliography{References}
\end{document}